\documentclass[aps,prd,onecolumn,groupedaddress,showpacs,nofootinbib,amssymb]{revtex4}
\usepackage[T1]{fontenc}
\usepackage[latin1]{inputenc}
\usepackage{graphicx}
\usepackage[english]{babel}
\usepackage{amsmath}
\usepackage{amssymb}
\usepackage{amsfonts}

\begin{document}

\def\pp{{\, \mid \hskip -1.5mm =}}
\def\cL{{\cal L}}
\def\be{\begin{equation}}
\def\ee{\end{equation}}
\def\bea{\begin{eqnarray}}
\def\eea{\end{eqnarray}}
\def\tr{{\rm tr}\, }
\def\nn{\nonumber \\}
\def\e{\mathrm{e}}

\title{Reconstruction of k-essence model}

\author{Jiro Matsumoto and Shin'ichi Nojiri}

\affiliation{
Department of Physics, Nagoya University, Nagoya 464-8602, Japan
}

\begin{abstract}

We explicitly construct the k-essence models which reproduce the arbitrary 
FRW cosmology, that is, the arbitrary time-development of the scale factor or the Hubble rate. 
The k-essence model includes scalar quintessence model, 
tachyon dark energy model, ghost condensation model as special cases. 
Explicit formulas of the reconstruction are given. 

First we consider the case that the action only contains 
the kinetic term. 
In this case, we find that the model reproducing the development of the universe 
in the exact $\Lambda$CDM model 
cannot be constructed although there is a model which reproduces the development 
infinitely closing to $\Lambda$CDM model. 
We find, however, the solution is not stable. 

Another is more general case including potential etc., where we find that there appear 
infinite number of arbitrary functions of the scalar fields, which are redundant 
to the time development of the scale factor. 
By adjusting one of the redundant functions, however, we can obtain the model 
where any solution we need could be stable. 

\end{abstract}

\pacs{95.36.+x, 98.80.Cq, 04.50.Kd, 11.10.Kk, 11.25.-w}

\maketitle

\section{Introduction}

Several observations tell that the expansion of the present universe is 
accelerating \cite{WMAP1,Komatsu:2008hk,SN1}. 
In order to explain the acceleration, many kinds of models have been proposed. 
So-called k-essence model \cite{Chiba:1999ka,ArmendarizPicon:2000dh,ArmendarizPicon:2000ah} 
is one of these models. The k-essence model is derived 
from k-inflation model \cite{ArmendarizPicon:1999rj,Garriga:1999vw}. 
It is possible to regard the tachyon dark energy 
model \cite{Sen:2002nu,Sen:2002an,Gibbons:2002md,Bagla:2002yn}, 
ghost condensation model \cite{ArkaniHamed:2003uy,ArkaniHamed:2003uz}, 
and scalar field quintessence model \cite{Peebles:1987ek,Ratra:1987rm,Chiba:1997ej,Zlatev:1998tr}
as variations of the k-essence model. 

In this paper we consider the reconstruction of k-essence model. 
We explicitly construct the k-essence models which reproduce the arbitrary 
FRW cosmology, that is, the arbitrary time-development of the scale factor or the Hubble rate. 
For general reconstruction, see \cite{Nojiri:2006be,reconstruction}.
We consider two cases: One is the case that the action only contains 
the kinetic term and another is more general case including potential etc. 
In the former case, we find that the exact $\Lambda$CDM model cannot be 
constructed although there is a model infinitely closing to $\Lambda$CDM model. 
In the model, however, the solution corresponding to $\Lambda$CDM model is 
unfortunately not stable. In the latter case, we find that there appear 
infinite number of redundant functions of the scalar fields, which are not directly 
related with the time development of the scale factor. 
By adjusting one of them, however, 
we can obtain the model where the solution we need could be stable. 
At present, the roles of the other functions are not clear but we may obtain 
a model satisfying other constraints from cosmology by adjusting these functions. 

\section{Pure Kinetic Model}

We may start with the following action:
\be
\label{M1}
S = \int d^4x \sqrt{-g}\left(\frac{R}{2\kappa ^2} 
 -K(q(\phi)\partial ^\mu \phi \partial _\mu \phi)\right)\, ,
\ee
where $K$ and $q$ are adequate functions. In the FRW background
\be
\label{M2}
ds^2 = -dt^2 + a(t)^2 \sum _{i=1,2,3} (dx^i)^2\, ,
\ee
we may assume the scalar field $\phi$ only depends on the time coordinate $t$. 
Since the redefinition of $\phi$ can be absorbed into the redefinition of $q(\phi)$, 
we further identify the scalar field $\phi$ with the time coordinate $t$. 
Then we obtain the following FRW equation: 
\be
\label{M3}
\frac{3}{\kappa ^2}H^2 = K(-q(t)) + 2K^ {\prime} (-q(t))q(t)\, ,
\ee
and the equation given by the variation of $\phi$, we obtain
\be
\label{M4}
0 = 2K ^{\prime \prime}(-q(t))q(t)q^{\prime}(t) - (6Hq(t) + q^{\prime}(t))K^{\prime}(-q(t))\, .
\ee
Here $H$ is the Hubble rate defined by $H = \dot a (t)/a(t)$. 
Eq. (\ref{M4}) can be integrated as
\be
\label{M5}
a(t)^6 = \frac{1}{q(t)} \left(\frac{K_0}{K^{\prime}(-q(t))}\right)^2\, .
\ee
By differentiating (\ref{M3}) with respect to $t$ and using (\ref{M5}), we find
\be
\label{M6}
\frac{6}{\kappa ^2} \dot H = -6 K_0 a(t)^{-3} \sqrt{q(t)}\, ,
\ee
which gives 
\be
\label{M7}
q(t) = \frac{a(t)^6 (H^{\prime} (t))^2}{\kappa ^4 K_0 ^2}\, .
\ee
Eq. (\ref{M7}) gives the explicit form of $q(t)$ 
and can be solved with respect to $t$ by using an adequate function $f$ as
\be
\label{M8}
t = f(q)\, .
\ee
Then (\ref{M5}) gives an explicit form of $K^{\prime}(-q)$ as 
\be
\label{M9}
K^{\prime} (-q) = \frac{K_0}{\sqrt{q}} a(f(q))^{-3}\, .
\ee
Then for the arbitrary time-development of $a$ or $H$ given by $a = a(t)$, 
Eqs. (\ref{M7}) and (\ref{M9}) give the forms of the functions $K$
and $q$ realizing the time-development. 
We should note, however, $\dot H $ cannot change its sign as we can see from (\ref{M6}). 
Then the transition between non-phantom phase ($ \dot H < 0$ ) 
and phantom phase ( $ \dot H > 0$ ) does not occur.

As an example, we consider the model corresponding to the Einstein gravity where the universe is filled 
with only one kind of perfect fluid whose equation of state (EoS) parameter $w$ is constant. 
In the universe, the scale factor $a$ and the Hubble rate $H$ behaves as
\be
\label{M10}
a(t) = a_0 t ^{\frac {2}{3(1+w)}}, \quad H = \frac {\frac{2}{3(1 + w)}}{t}\, ,
\ee
with a constant $a_0$. Then Eq. (\ref{M7}) gives 
\be
\label{M11}
q(t) = q_0 t^{\frac {4w}{1+w}}\, , \quad 
q_0 \equiv \frac {a_0 ^6}{\kappa ^4 K_0 ^2}\left( \frac {2}{3(1 + w)} \right)^2\, ,
\ee
and Eq. (\ref{M9}) has the following form:
\be
\label{M12}
K^{\prime} (-q) = K_0 q_0 ^{- \frac {1}{2w}} a_0 ^{-3} q^{\frac {1}{2w} - \frac{1}{2}}\, .
\ee
We can integrate (\ref{M12}) as
\be
\label{M13}
K(-q) = K_1 - \frac {2w}{w+1} K_0 q_0 ^{- \frac {1}{2w}}a_0 ^{-3} q^{\frac {1+w}{2w}}\, ,
\ee
where $K_1$ is a constant of the integration. By using (\ref{M11}), we find
\be
\label{M14}
K(-q) = K_1 - \frac {4w}{3(1+w)^2 \kappa ^2 t^2}, 
\quad 2K^{\prime}(-q)q = \frac {4}{3(1+w) \kappa ^2 t^2}\, .
\ee
Then we can check that Eq. (\ref{M3}) is satisfied if $K_1 = 0$ 
and we find the Einstein gravity where the universe filled with only one kind of perfect fluid 
can be reproduced by the model (\ref{M1}) if we choose $q(t)$ and $K(-q)$ by (\ref{M11}) 
and (\ref{M13}) with $K_1 = 0$.

We now extend the model (\ref{M1}) to include the matters with constant EoS parameters $w_i$. 
Then since the energy density of the matters is given by $\sum _i \rho_{0i}a^{-3(1+w_i)}$ with 
constants $\rho_{0i}$, the FRW equation (\ref{M3}) is modified as
\be
\label{M15}
\frac{3}{\kappa ^2}H^2 = K(-q(t)) + 2K^{\prime}(-q(t))q(t) + \sum _i \rho_{0i}a^{-3(1+w_i)}\, .
\ee
Eq. (\ref{M4}) given by the variation of $\phi$ is not changed and 
we obtain (\ref{M5}) again. On the other hand, Eq. (\ref{M6}) is changed to
\be
\label{M16}
\frac {6}{\kappa ^2} \dot H = -6 K_0 a(t)^{-3} \sqrt{q(t)} 
 - \sum _i 3(1+w_i) \rho _{0i} a^{-3(1+w_i)}\, ,
\ee
which gives
\be
\label{M17}
q(t) = \frac {a(t)^6 (H^{\prime}(t) + \frac{\kappa ^2}{2} 
\sum _i (1+w_i) \rho _{0i} a^{-3(1+w_i)})^2}{\kappa ^4 K_0 ^2}\, .
\ee
Eq. (\ref{M17}) gives the explicit form of $q(t)$ as in (\ref{M7}) 
and can be solved with respect to $t$ by using an adequate function $f$ 
as in (\ref{M8}), again. Then Eq.(\ref{M5}) gives an explicit form of 
$K^{\prime} (-q)$ by (\ref{M9}).


If we substitute the evolution of scale factor $a(t)$,
\be
\label{M1_1}
a(t) = A \sinh ^{\frac {2}{3}} [\alpha t]\, , 
\ee
which corresponds to the $\Lambda$CDM model in the Einstein gravity, 
to Eqs. (\ref{M17}) and (\ref{M12}), then these equations give constant $q$ and $K$. 
This result seems to be very trivial since $K$ is nothing but the cosmological constant.  
Then we now consider the evolution of scale factor which differs a little bit from that 
of $\Lambda$CDM model in the Einstein gravity.
For this purpose, we slightly modify (\ref{M1_1}) as
\be
\label{M1_2}
a(t) = A \exp \left[\ln (\sinh^{\frac {2}{3}} \left[\alpha t\right]) + \epsilon f(t)\right],
\ee
where $A = ( \frac {\rho _{0m}}{\Lambda} )^{\frac {1}{3}}$ and $\alpha = \frac {\kappa}{2} \sqrt{3 \Lambda} $ 
are positive constants which reproduce the universe evolution in $\Lambda$CDM model, 
$f$ is an arbitrary function of $t$, and $\epsilon$ is a small constant. 
Then Eq. (\ref{M17}) gives
\be
\label{M1_3}
q(t) \sim \frac {\epsilon ^2}{\kappa ^4 K_0 ^2} \left( \frac {9}{4} \kappa ^4 \rho _{0m}^2 f^2
 -3 \kappa ^2 \frac {\rho _{0m} ^2}{\Lambda} \sinh ^2 [\alpha t] ff^{\prime \prime} 
+ \frac {\rho _{0m}^2}{\Lambda ^2} \sinh ^4 [\alpha t] (f^{\prime \prime})^2 \right)\, ,
\ee
where we neglected the higher order terms of $\epsilon$. 
As an examples, we may consider the following function as $f$,
\be
\label{M1_4}
f(t) = \ln \left( \sinh^{\frac {2}{3}}[\alpha t] \right) \ \mbox{s.t.} 
\ a(t) = A(\sinh ^{\frac {2}{3}}[\alpha t])^{1+\epsilon}\, .
\ee
By substituting (\ref{M1_4}) to (\ref{M1_3}), we find 
\be
\label{M1_8}
t \sim \frac {1}{\alpha} \sinh ^{-1} \left[ \exp \left[- \frac {1}{2} + \frac {K_0}{\rho _{0m} \epsilon} \sqrt{q} 
\right] \right]\, .
\ee
Then by further substituting (\ref{M1_8}) into Eq. (\ref{M5}), we obtain
\be
\label{M1_9}
K(-q) \sim \frac {K_0 \Lambda}{\rho _{0m}} \int ^q dq^{\prime} q^{\prime - \frac {1}{2}} 
\left( \exp \left[- \frac {1}{2} + \frac {K_0}{\rho_{0m} \epsilon} \sqrt{q^{\prime}} \right] 
\right)^{-2(1+ \epsilon)}
= \frac{\epsilon \Lambda \e^{1+\epsilon}}{1+\epsilon} 
\left( C - \exp \left[ -\frac{2K_0 (1+\epsilon)\sqrt{q}}{\rho_{0m} \epsilon} \right] \right)\, .
\ee
Here $C$ is a constant of the integration. 
Then we find the existence of the functions $q$ and $K$, which give $a(t)$ very close to that of $\Lambda$CDM model.


We now investigate the stability of the solution.  
So we now start with the two FRW equations and the field equation obtained from Eq. (\ref{M1}), 
\bea
\label{M2_1}
&& \frac {3}{\kappa ^2}H^2 = K(-q \dot \phi ^2) + 2K^{\prime}(-q \dot \phi ^2)q \dot \phi ^2\, , \\
\label{M2_2}
&& \frac {1}{2 \kappa ^2}(3H^2 + 2 \dot H) = \frac {1}{2}K(-q \dot \phi ^2)\, , \\
\label{M2_3}
&& \frac {d}{dt}(2a^3 q^{\frac {1}{2}} \dot \phi K^{\prime}(-q \dot \phi ^2)) = 0\, .
\eea
 From Eqs. (\ref{M2_1}) and (\ref{M2_2}), we obtain
\be
\label{M2_4}
K^{\prime} (-Q) Q = - \frac {\dot H}{\kappa ^2}\, .
\ee
Here $Q \equiv q \dot \phi ^2$. Inserting this equation to Eq. (\ref{M2_3}), we find
\be
\label{M2_5}
\frac {d}{dt} \Big(2a^3 Q^{- \frac {1}{2}} \frac {\dot H}{\kappa ^2} \Big) = 0\, ,
\ee
which can be integrated and gives
\be
\label{M2_6}
\frac{1}{\kappa ^2}a^3 Q^{- \frac {1}{2}} \dot H \equiv K_0\, .
\ee
Here $K_0$ is a constant of the integration. 
We now write the form of $a(t)$ as $a(t) = a_0 \e^{h(t)}$ and consider the perturbation of the function $h(t)$: 
$h(t) = h_0(t) + \delta h$.
Then from Eq. (\ref{M2_6}), we find the variation of $Q$ is given by
\be
\label{M2_7}
\delta Q = \frac {a_0 ^6}{\kappa ^4 K_0 ^2} \e^{6h_0}
\left(6 \delta h \ddot h_0 ^2 +2 \ddot h_0 \delta \ddot h \right)\, .
\ee
By combining Eq.(\ref{M2_7}) and Eq. (\ref{M2_4}), we obtain
\be
\label{M2_8}
K^{\prime \prime}(-Q_0)Q_0 ^2 
= - \frac {\ddot h_0}{\kappa ^2} 
+ \frac {\dddot h_0}{\kappa ^2 \left( 6 \dot h_0 + \frac {2 \dddot h_0}{\ddot h_0} \right)}\, .
\ee
Here $Q_0$ is defined by substituting $h(t)=h_0(t)$ into (\ref{M2_6}): 
$Q_0 \equiv a_0^6 \e^{6h_0(t)} \ddot h_0(t)^2 / K_0^2 \kappa^4$. 
Substituting Eqs. (\ref{M2_4}), (\ref{M2_7}), and (\ref{M2_8}) into the equation,
$ \frac {6}{\kappa ^2} \dot h_0 \delta \dot h = \Big( K^{\prime}(-Q_0)-2K^{\prime \prime }(-Q_0)Q_0 \Big) \delta Q $,
which is given by the variation of Eq. (\ref{M2_1}), we obtain a rather simple differential equation for $\delta h(t)$,
\be
\label{M2_9}
\delta \ddot h -  \frac {3 \dot h_0 \ddot h_0 + \dddot h_0}{\ddot h_0}  \delta \dot h + 3 \ddot h_0 \delta h = 0\, .
\ee
The above equation only contains $h_0$ and its derivatives. Then if we have $h_0$, even if we do not know the 
explicit forms of $q$ and $K$, we can find the stability of the solution. 
In case that solution $h_0$ is given by (\ref{M1_2}), by choosing $a_0=A$, 
\be
\label{N1}
h_0 = \ln (\sinh^{\frac {2}{3}} \left[\alpha t\right]) + \epsilon f(t)\, ,
\ee
Eq. (\ref{M2_9}) has the following form:
\be
\label{N2}
\delta \ddot h  - \frac {2 \alpha ^2}{\sinh ^2 [\alpha t]} \delta h + \mathcal{O}(\epsilon) = 0\, .
\ee
Since $\frac {2 \alpha ^2}{\sinh ^2 [\alpha t]}>0$, the solution (\ref{N1}) is unstable. 
Then unfortunately the solution which behaves close to the $\Lambda$CDM model is not stable. 

By defining $x\equiv \delta h$ and $y= \delta\dot h$, Eq.(\ref{M2_9}) can be rewritten as
\be
\label{N3}
\frac{d}{dt} \left(\begin{array}{c}
x \\ y \end{array} \right) 
= \left( \begin{array}{cc}
0 & 1 \\
- 3 \ddot h_0 & \frac {3 \dot h_0 \ddot h_0 + \dddot h_0}{\ddot h_0}  
\end{array} \right) 
\left(\begin{array}{c}
x \\ y \end{array} \right) \, .
\ee
In order that the solution could be stable, the real part of all the eigenvalues of the 2$\times$2 matrix 
in (\ref{N3}) should be negative:
\be
\label{N4}
\Re \lambda_\pm < 0\, ,\quad 
\lambda_\pm \equiv \frac{1}{2}\left[ \frac {3 \dot h_0 \ddot h_0 + \dddot h_0}{\ddot h_0} \pm 
\sqrt{ \left( \frac {3 \dot h_0 \ddot h_0 + \dddot h_0}{\ddot h_0} \right)^2 - 12 \ddot h_0} \right] \, ,
\ee
which requires $\ddot h_0>0$. 
This tells that the solution corresponding to non-phantom era, like quintessence era or 
matter dominance era, is not stable. 

We now investigate the stability when we include the matters as in (\ref{M15}). 
Then the equations corresponding to (\ref{M2_1}) - (\ref{M2_3}) have the following form:
\bea
\label{M3_1}
&& \frac{3}{\kappa ^2}H^2 = K(-Q) + 2K^{\prime}(-Q)Q + \sum _i \rho _{0i} a^{-3(1+w_i)}\, ,\\
\label{M3_2}
&& \frac{1}{2 \kappa ^2}(3H^2 2 \dot H) = \frac{1}{2} K(-Q) - \frac{1}{2} \sum _i w_i \rho _{0i} a^{-3(1+w_i)}\, ,\\
\label{M3_3}
&& \frac{d}{dt}(2a^3Q^{\frac{1}{2}}K^{\prime}(-Q))=0\, .
\eea
Then by considering the perturbation $h \rightarrow h_0 + \delta h$, we find
\bea
\label{M3_4}
0 &=& \delta \ddot h -b \delta \dot h -c \delta h\, ,\\
\label{M3_5}
b &\equiv & 3 \frac{\frac{1}{\kappa ^2}(3 \dot h_0 \ddot h_0 + \dddot h_0) 
 - \frac{3}{2}\sum _i w_i (1+w_i) \rho _{0i} a_0^{-3(1+w_i)}\e^{-3(1+w_i)h_0}\dot h_0}
{\frac {3}{\kappa^2} \ddot h_0 + \frac{3}{2} \sum _i (1+w_i) \rho _{0i} a_0^{-3(1+w_i)}\e^{-3(1+w_i)h_0}}\, ,\\
\label{M3_6}
c &\equiv & -\frac{\kappa^2}{2}\frac{\frac{18}{\kappa ^4}\ddot h_0^2 -3 \sum_i (1+w_i) 
\rho _{0i}a_0^{-3(1+w_i)} \e^{-3(1+w_i)h_0}\frac{1}{\kappa^2}
\left(6w_i \ddot h_0 + (1+w_i)\frac{\dddot h_0}{\dot h_0} \right)}
{\frac{3}{\kappa^2} \ddot h_0 + \frac{3}{2} \sum _i (1+w_i) \rho _{0i} a_0^{-3(1+w_i)} \e^{-3(1+w_i)h_0}} \nn
&& -\frac{\kappa^2}{2}\frac{\frac{9}{2}\sum_{i,j}w_i(1+w_i)(1+w_j)\rho_{0i}\rho_{0j}
a_0^{-3(2+w_i+w_j)} \e^{-3(2+w_i+w_j)h_0}}{\frac {3}{\kappa^2} 
\ddot h_0 + \frac{3}{2} \sum _i (1+w_i) \rho _{0i} a_0^{-3(1+w_i)} \e^{-3(1+w_i)h_0}}\, .
\eea
The above expression is rather complicated but in principle, we can check the stability by using the above 
equations. As long as we check numerically, the solution (\ref{N1}) which behaves as almost $\Lambda$CDM solution in 
the Einstein gravity seems to be unstable. 

\section{More general models}

Since in the model (\ref{M1}), the solution which reproduces almost $\Lambda$CDM is unstable, 
we consider a more general model, whose action is given by
\be
\label{KK1}
S=  \int d^4 x \sqrt{-g} \left( \frac{R}{2\kappa^2} - K \left( 
\phi, X \right) + L_\mathrm{matter}\right)\, ,\quad X \equiv \partial^\mu \phi \partial_\mu \phi \, .
\ee
Then the FRW equations are given by
\be
\label{KK2}
\frac{3}{\kappa^2} H^2 = 2 X \frac{\partial K\left( \phi, X \right)}{\partial X} 
 - K\left( \phi, X \right) + \rho_\mathrm{matter}\, ,\quad 
 - \frac{1}{\kappa^2}\left(2 \dot H + 3 H^2 \right) 
= K\left( \phi, X \right) + p_\mathrm{matter}(t)\, .
\ee
As in (\ref{M15}), we consider the matters which have constant EoS parameters $w_i$. 
We now choose $\phi=t$, again, then we can rewrite the equations in (\ref{KK2}) in the following form
\be
\label{KK4}
K\left( t, -1 \right) = - \frac{1}{\kappa^2}\left(2 \dot H + 3 H^2 \right) 
 - \sum_i w_i\rho_{0i} a^{-3(1+w_i)}\, ,\quad 
\left. \frac{\partial K\left( \phi, X \right)}{\partial X}\right|_{X=-1} 
=  \frac{1}{\kappa^2} \dot H  + \frac{1}{2}\sum_i \left(1+w_i\right) \rho_{0i} a^{-3(1+w_i)}\, .
\ee
By using the appropriate function $g(\phi)$, if we choose 
\bea
\label{KK5}
K(\phi,X) &=& - \frac{1}{\kappa^2}\left(2 g''(\phi) + 3 g'(\phi)^2 \right)
 - \sum_i w_i\rho_{0i} a_0^{-3(1+w_i)}\e^{-3(1+w_i)g(\phi)} \nn 
&& + \left(X+1\right) \left\{  \frac{1}{\kappa^2} g''(\phi)  + \frac{1}{2}\sum_i \left(1+w_i\right) 
\rho_{0i} a_0^{-3(1+w_i)}\e^{-3(1+w_i)g(\phi)}\right\}
+ \sum_{n=2}^\infty \left(X+1\right)^n K^{(n)} (\phi)
\, ,
\eea
we find the following solution for the FRW equations (\ref{KK2}), 
\be
\label{KK6}
H= g'(t) \quad \left(a = a_0 \e^{g(t)} \right)\, .
\ee
Here $K^{(n)}(\phi)$ with $n=2,3,\cdots$ can be arbitrary functions. 
The case that $K^{(n)}(\phi)$'s with $n=2,3,\cdots$ vanish was studied in 
\cite{Nojiri:2005pu,Capozziello:2005tf} and the instability was investigated. 

We now investigate the stability of the equations without matter.
 From Eqs. (\ref{KK2}), we can derive the following equation which does not contain 
the variable $g^{\prime \prime}$,
\be
\label{KK7}
3 \frac{1-y^2}{1+X}X= -\frac{\dot H}{H^2}+ \frac{\kappa ^2}{H^2} 
\sum ^{\infty}_{n=2} \left( (n-1)X-n-1 \right)X(X+1)^{n-2}K^{(n)}(\phi)\, ,
\ee
where $y=\frac{g^{\prime}}{H}$. 
Using Eq. (\ref{KK7}), we can rewrite $dy/dN = \left(1/H\right)dy/dt$ 
($N$ is called e-foldings and the scale factor is given in terms of $N$ as $a\propto \e^N$) 
in the form which does not contain $g$:
\be
\label{KK8}
\frac{dy}{dN} = 3X \frac{1-y^2}{1+X} \left(\frac{\dot \phi}{X} +y \right) 
 - \frac{\kappa^2}{H^2} \sum ^{\infty}_{n=2} \left[ ( \dot \phi +yX) \left( (n-1)X -n-1 \right) 
+ \dot \phi n(X+1) \right](X+1)^{n-2}K^{(n)}(\phi)\, .
\ee
We now consider the perturbation from a solution $\phi=t$ 
by putting $\phi = t + \delta \phi$ in (\ref{KK8}). Then we obtain
\be
\frac{d \delta \dot \phi}{dN} = \left[ -3- \frac{g^{\prime \prime}}{g^{\prime 2}} - \frac{d}{dN} 
\left\{ \frac{ \kappa ^2}{6g^{\prime 2}}(8K^{(2)}- \frac{2}{\kappa ^2}g^{\prime \prime}) \right\} \right] 
\delta \dot \phi\, .
\ee
If the quantity inside $[\ ]$ is negative, the fluctuation $\delta\dot\phi$ becomes exponentially smaller 
with time and therefore the solution becomes stable. 
Note that the stability is determined only in terms of $K^{(2)}$ and does not depend on other $K^{(n)}$ ($n\neq 2$). 
Then if we choose $K^{(2)}$ properly, the solution corresponding to arbitrary development of the universe becomes 
stable. 

We now investigate the stability when we include the matter. Then the equation corresponding to Eq.(\ref{KK7}) 
has the following form:
\bea
\label{KK18}
3 \frac{1-y^2}{1+X}X &=& -\frac{\dot H}{H^2} + \frac{\kappa ^2}{H^2} \sum ^{\infty}_{n=2} 
\left( (n-1)X-n-1\bigg)X(X+1)^{n-2}K^{(n)}(\phi) \right. \nn 
&& + \frac{\kappa^2}{H^2} \frac{X-1}{2(X+1)} \rho_\mathrm{matter} 
 - \frac{\kappa^2}{2H^2}p_\mathrm{matter}-\frac{\kappa^2}{H^2}\frac{X}{X+1} 
\sum_i \rho _{0i} a_0 ^{-3(1+w_i)} \e^{-3(1+w_i)g(\phi)}\, .
\eea
Then we find
\bea
\label{KK19}
\frac{dy}{dN} &=& 3X \frac{1-y^2}{1+X} \left(\frac{\dot \phi}{X} +y \right) 
 - \frac{\kappa^2}{H^2} \sum^{\infty}_{n=2} \left[ ( \dot \phi +yX) \left( (n-1)X -n-1 \right) 
+ \dot \phi n(X+1) \right](X+1)^{n-2}K^{(n)}(\phi) \nonumber \\
&& + \frac{\kappa ^2}{2H^2 X} \left( -\frac{X-1}{X+1}\left( \dot \phi + yX \right) - \dot \phi \right) 
\rho_\mathrm{matter} + \frac{\kappa ^2}{2H^2}yp_\mathrm{matter} \nonumber \\
&& + \frac{\kappa ^2}{2H^2} \sum_i \left( \left( \dot \phi +yX \right) \frac{2}{X+1} - \dot \phi (1+w_i) \right) 
\rho _{0i} a_0^{-3(1+w_i)} \e^{-3(1+w_i)g(\phi)}\, .
\eea
When we include the matter, the situation becomes a little bit complicated since not only $H$ but the scale factor 
$a$ appears in the equation. Then we need equation to describe the time development of $a$. 
By defining $\delta\lambda$ for convenience as
\be
\label{KK24}
\delta \lambda \equiv 3 \sum _i (1+w_i) \rho _{0i} a(t)^{-3(1+w_i)} \frac{\kappa ^2}{6 g^{\prime 2}} 
\left( \frac{\delta a}{a} -g ^{\prime} \delta \phi \right)\, .
\ee
we obtain the following equations,
\be
\label{KK29}
\left. \frac{d}{dN} \left(
\begin{array}{c}
 \delta \dot \phi \\
 \delta \lambda
\end{array}
\right)
\right| _{\phi =t, H=g^{\prime}(t)} =
 \left(
\begin{array}{cc}
A & B \\
C & D
\end{array}
\right)
\left(
\begin{array}{c}
 \delta \dot \phi \\
 \delta \lambda
\end{array}
\right) \, .
\ee
Here
\bea
\label{KK30}
A &\equiv& -3 + \frac{g ^{\prime \prime}}{g^{\prime 2}} + \frac{\kappa ^2}{2 g^{\prime 2}} 
\sum _i (1+w_i) \rho _{0i} a(t)^{-3(1+w_i)} 
 - \frac{d}{dN} \ln \left\{ 8K^{(2)} - \frac{2}{\kappa ^2}g^{\prime \prime} - \sum _i (1+w_i) 
\rho _{0i} a(t)^{-3(1+w_i)} \right\} \, ,\nonumber \\
B &\equiv& 3 - \frac{24 K^{(2)}}{8K^{(2)} - \frac{2}{\kappa^2} g^{\prime \prime} 
 - \sum _i (1+w_i) \rho _{0i} a(t)^{-3(1+w_i)}} \, , \nonumber \\
C &\equiv& 3 \sum _i (1+w_i) \rho _{0i} a(t)^{-3(1+w_i)} \left( \frac{\kappa ^2}{6 g^{\prime 2}} \right)^2 
\left\{ 8K^{(2)}- \frac{6g^{\prime 2}}{\kappa ^2} - \frac{2}{\kappa ^2}g^{\prime \prime}
 - \sum _i (1+w_i) \rho _{0i} a(t)^{-3(1+w_i)}  \right\} \, , \nonumber \\
D &\equiv& \frac{d}{dN} \ln \left\{ \frac{\kappa ^2}{2 g^{\prime 2}} 
\sum _i (1+w_i) \rho _{0i} a(t)^{-3(1+w_i)} \right\}
 - \frac{\kappa ^2}{2g^{\prime 2}} \sum _i (1+w_i) \rho _{0i} a(t)^{-3(1+w_i)}\, .
\eea
Generally, 2$\times$2 matrix must have negative trace and positive determinant in order that 
two eigenvalues could be negative since the two eigenvalues are given by 
$\frac{1}{2}\{ \tr M \pm \sqrt{(\tr M)^2-4(\det M)} \}$. 
So we just need to calculate the determinant and the trace of the matrix for investigating the stability 
of the fixed point $\phi =t$, $H=g^{\prime}(t)$. 
Since the expressions in (\ref{KK30}) are very complicated, we consider the case that the solution is given 
by the behavior mimicing $\Lambda$CDM solution (\ref{M1_1}) in the Einstein gravity 
and the matter contents are given in the present universe. 
Then we find
\bea
\label{KK31}
a(t) &\sim& A \sinh ^{\frac{2}{3}} [\alpha t]\, , \\
\label{KK32}
g^{\prime}(t) &\sim& \frac{2}{3} \alpha \coth [\alpha t]\, , \\
\label{KK33}
\frac{\kappa ^2}{\alpha ^2} \sum _i (1+w_i) \rho _{0i} a(t)^{-3(1+w_i)} 
&\sim& \frac{2}{3} \frac{1}{\sinh ^2 [\alpha t]}\, , \\
\label{KK34}
\frac{\kappa ^2}{\alpha ^2} \sum _i w_i (1+w_i) \rho _{0i} a(t)^{-3(1+w_i)} 
&\sim& \frac{4}{9} \times 1.86 \times 10^{-4} \frac{1}{\sinh ^{\frac{8}{3}}[\alpha t]}\, , 
\eea  
where $A \equiv (\rho _{m0}/ \rho _{\Lambda})^{\frac{1}{3}}$ and $\alpha \equiv \kappa \sqrt{3 \rho _{\Lambda}} /2$. 
Note that in (\ref{KK31}), (\ref{KK32}), and (\ref{KK33}), we neglect the contribution from radiation, and 
in (\ref{KK34}), there only appears the contribution from radiation. 
Therefore the expressions in (\ref{KK31}) - (\ref{KK34}) could be valid at least when $t \geq 10^9$ years. 
By using (\ref{KK31}) - (\ref{KK34}), we find the following expressions of the determinant and trace of the matrix 
in Eq.(\ref{KK29}):
\bea
\label{KK35}
\tr \left(
\begin{array}{cc}
A & B \\
C & D
\end{array}
\right) &\sim&
 -6 + \frac{3}{2} \frac{1}{\cosh ^2 [\alpha t]} - \frac{8 \frac{\kappa ^2}{\alpha ^3}K^{(2) \prime}
 - \frac{4}{3} \frac{\cosh [\alpha t]}{\sinh ^3 [\alpha t]} }
{ \frac{2}{3} \coth [\alpha t] \Big( 8 \frac{\kappa ^2}{\alpha ^2}K^{(2)} 
+ \frac{2}{3} \frac{1}{\sinh ^2 [\alpha t]} \Big)}\, , \\
\label{KK36}
\det \left(
\begin{array}{cc}
A & B \\
C & D
\end{array}
\right) &\sim& 
\left\{ 8 \frac{\kappa ^2}{\alpha ^2}K^{(2)} + \frac{2}{3} \frac{1}{\sinh ^2 [\alpha t]}  \right\} ^{-1}
\left[ \left( -27 \frac{\sinh [\alpha t]}{\cosh ^3 [\alpha t]} + 36 \tanh [\alpha t]  \right) 
\frac{\kappa ^2}{\alpha ^3}K^{(2) \prime} \right. \nonumber \\
&& + \left( 72-36 \frac{1}{\cosh ^2 [\alpha t]} -18 \frac{1}{\cosh ^4 [\alpha t]} 
 -12 \times 1.86 \times 10^{-4} \frac{1}{\sinh ^{\frac{8}{3}}[\alpha t]} \right) 
\frac{\kappa ^2}{\alpha ^2}K^{(2)} \nonumber \\
&& - 6 \times 1.86 \times 10^{-4} \frac{1}{\cosh ^2[\alpha t] \sinh ^{\frac{8}{3}}[\alpha t]}
 - \frac{3}{2} \frac{1}{\cosh ^4 [\alpha t] \sinh ^2 [\alpha t]} \nonumber \\
&& \left. + 3 \frac{1}{\sinh ^2 [\alpha t] \cosh ^2 [ \alpha t]} + 4 \times 1.86 \times 10^{-4} 
\frac{1}{\sinh ^{\frac{8}{3}}[\alpha t]}  \right]\, .
\eea 
Note that $1 < \cosh [ \alpha t] \leq 2$ and $0 < \sinh [ \alpha t] \leq 1.7$ in evolution of the universe, 
For example, if we consider the case that $ K^{(2)} $ is constant, then the trace of the matrix is always negative 
and the determinant is positive when $ K^{(2)} \geq 0 $ since $72 \frac{\kappa ^2}{\alpha ^2}K^{(2)}$ 
and $3/( \sinh ^2 [\alpha t] \cosh ^2 [\alpha t])$ are dominant terms in Eq.(\ref{KK36}). 
Therefore even if $K^{(2)}$ is constant, the fixed point solution mimicing $\Lambda$CDM solution in the Einstein gravity 
becomes stable as long as $K^{(2)} \geq 0$. 

\section{Tachyon model as a generalized k-essence model}

We now consider the tachyon model \cite{Sen:2002nu,Sen:2002an,Gibbons:2002md,Bagla:2002yn} as 
a variation of the generalized k-essence model. 

The action of the tachyon model is given by
\be
\label{T37}
S_\mathrm{BI} = \tau _3 \int d^4 x \sqrt{-g} V(T) \sqrt{1+ l_s ^2 f(T) \partial_\mu T \partial ^ \mu T}\, .
\ee 
Here $l_s$ is the string scale and $\tau_3$ is defined by using the string coupling constant $g_s$ as 
$\tau_3 \equiv 1/(8 \pi ^3  g_s l_s ^4)$. 
Since the scalar field $T$ is dimensionless, we define a new field $\phi$ by $\phi \equiv \kappa T$ 
and redefine the functions $V(T)$ and $f(T)$ as $V(T)= -v(\phi)/ \tau _3$ 
and $f(T)= \tilde{\omega} (\phi) \kappa ^2 / l_s ^2$. Then the action (\ref{T37}) can be rewritten as
\be
\label{T38}
S_\mathrm{BI} = -\int d^4 x \sqrt{-g} v(\phi) 
\sqrt{1+ \tilde{\omega} (\phi) \partial_ \mu \phi \partial ^ \mu \phi}\, .
\ee
We now consider the FRW background (\ref{M2}). Since the redefinition of the scalar field $\phi$ 
can be absorbed into the redefinition of $\tilde\omega(\phi)$, later we may identify $\phi$ with the 
cosmological time $t$. Then by expanding the square root in the action (\ref{T37}) by the power of 
$ 1 + \partial_ \mu \phi \partial ^ \mu \phi \sim 0$, we find, 
\bea
\label{T39}
\sqrt{1+ \tilde{\omega} (\phi) X} &=& \sqrt{1- \tilde{\omega}(\phi) + \tilde{\omega}(\phi)(X+1)} \nonumber \\
&=& \sqrt{1- \tilde{\omega}(\phi)} + \frac{\tilde{\omega}(\phi)}{2 \sqrt{1- \tilde{\omega}(\phi)}}(X+1) 
- \frac{\tilde{\omega}^2(\phi)}{8(1- \tilde{\omega}(\phi))^{\frac{3}{2}}}(X+1)^2 
+ \mathcal{O} \left( \left(X+1\right)^3 \right)\, ,
\eea
where $X \equiv \partial_ \mu \phi \partial ^ \mu \phi$. 
Then by comparing $S_\mathrm{BI}$ in (\ref{T38}) with Eq.(\ref{KK1}) and Eq.(\ref{KK5}), 
we can identify
\bea
\label{T40}
v(\phi) &=& \left\{- \frac{1}{\kappa^2}\left(2 g''(\phi) + 3 g'(\phi)^2 \right)
 - \sum_i w_i\rho_{0i} a_0^{-3(1+w_i)}\e^{-3(1+w_i)g(\phi)} \right\}^{\frac{1}{2}} \nonumber \\
&& \times \left\{ - \frac{3}{\kappa^2} g'(\phi)^2 
 + \sum_i\rho_{0i} a_0^{-3(1+w_i)}\e^{-3(1+w_i)g(\phi)}  \right\}^{\frac{1}{2}}\, , \\
\label{T41}
\tilde{\omega}(\phi) &=& \frac{ \frac{2}{\kappa^2} g''(\phi) 
 + \sum_i (1+w_i)\rho_{0i} a_0^{-3(1+w_i)}\e^{-3(1+w_i)g(\phi)}}{- \frac{3}{\kappa^2} g'(\phi)^2
 + \sum_i \rho_{0i} a_0^{-3(1+w_i)}\e^{-3(1+w_i)g(\phi)} }\, , \\
K^{(2)}(\phi) &=& - \frac{1}{8} \frac{ \tilde{\omega}^2(\phi)}{(1- \tilde{\omega}(\phi))^{\frac{3}{2}}} 
\label{T42}
v(\phi) \nonumber \\
&=& \frac{1}{8} \frac{ \left\{  \frac{2}{\kappa^2} g''(\phi) 
 + \sum_i (1+ w_i)\rho_{0i} a_0^{-3(1+w_i)}\e^{-3(1+w_i)g(\phi)} \right\} ^2 }
{ \frac{1}{\kappa^2}\left(2 g''(\phi) + 3 g'(\phi)^2 \right)
 + \sum_i w_i\rho_{0i} a_0^{-3(1+w_i)}\e^{-3(1+w_i)g(\phi)} }\, .
\eea
If we consider the case that the action $S_\mathrm{BI}$ reproduces the $\Lambda$CDM behavior in the Einstein gravity, 
by using Eqs.(\ref{KK31})-(\ref{KK34}), we find
\bea
\label{T43}
K^{(2)}(t) &\sim& \frac{\alpha ^2}{6 \kappa ^2} \frac{1}{-4 \sinh ^2 [\alpha t] 
+ 4 \cosh ^2 [\alpha t] \sinh ^2 [\alpha t] +1.86 \times 10^{-4} \sinh ^{\frac{4}{3}} [\alpha t]} > 0\, , \\
K^{(2) \prime}(t) &\sim& - \frac{\alpha ^3}{\kappa ^2} \frac{2 \cosh [\alpha t]}{9 \sinh ^3 [\alpha t] 
\left\{ -4+4 \cosh ^2 [\alpha t] + 1.86 \times 10^{-4} \frac{1}{ \sinh ^{\frac{2}{3}}[\alpha t]}  \right\} ^2} \nonumber \\
&& \times \left( -12+12 \cosh ^2 [\alpha t] +1.86 \times 10^{-4} \frac{1}{ \sinh ^{\frac{2}{3}} [\alpha t]} \right) < 0\, .
\eea 
Then we can find the stability of the fixed point corresponding to the $\Lambda$CDM model (\ref{M1_1}) 
by substituting 
the equations in (\ref{T43}) into Eq.(\ref{KK35}) and (\ref{KK36}). 
Since the expression is, however, very complicated, it is difficult to check the signs of 
the determinant and the trace analytically. Numerical calculation, however, tells that 
both of the determinant and the trace of the matrix are negative and therefore  
the fixed point solution corresponding to the $\Lambda$CDM model is unstable. 
So this model cannot reproduce the stable evolution of exact $\Lambda$CDM model.

\section{Summary}

In this paper, we consider the reconstruction of k-essence model which includes scalar quintessence model, 
tachyon dark energy model, ghost condensation model as special cases. 
Explicit formulas of the reconstruction were given. 
First we considered the case that the action only contains 
the kinetic term. 
In this case, we find that the model reproducing the development of the universe 
in the exact $\Lambda$CDM model 
cannot be constructed although there is a model which reproduce the development 
infinitely closing to that of $\Lambda$CDM model. 
We find, however, the solution is not stable. 

Another is more general case including potential etc., where we find that there appear 
infinite number of functions of the scalar fields, which are denoted by $K^{(n)}(\phi)$ ($n=2,3,4,\cdots$) 
and redundant to the time development of the scale factor. 
Although $K^{(2)}(\phi)$ is also redundant to the time development of the scale factor, 
we found that by adjusting $K^{(2)}(\phi)$, we can obtain the model where any solution we need 
could be stable. 
At present, the roles of the other functions $K^{(n)}(\phi)$ ($n=3,4,\cdots$) are not clear 
but we expect that we may obtain a model satisfying other constraints from cosmology 
by adjusting these functions. 

\section*{Acknowledgments}

We are grateful to S.~D.~Odintsov for very helpful discussions.
The work by S.N. is supported by Global
COE Program of Nagoya University provided by the Japan Society
for the Promotion of Science (G07).



\begin{thebibliography}{99}

\bibitem{WMAP1}
D.~N.~Spergel {\it et al.} [WMAP Collaboration],
Astrophys.\ J.\ Suppl.\  {\bf 148}, 175 (2003);\
H.~V.~Peiris {\it et al.}  [WMAP Collaboration],
\textit{ibid}. {\bf 148}, 213 (2003);\
D.~N.~Spergel {\it et al.}  [WMAP Collaboration],
\textit{ibid}. {\bf 170}, 377 (2007).

\bibitem{Komatsu:2008hk}
E.~Komatsu {\it et al.}  [WMAP Collaboration],
Astrophys.\ J.\ Suppl.\  {\bf 180}, 330 (2009).

\bibitem{SN1}
S.~Perlmutter {\it et al.}  [Supernova Cosmology Project Collaboration],
Astrophys.\ J.\  {\bf 517}, 565 (1999);\
A.~G.~Riess {\it et al.}  [Supernova Search Team Collaboration],
Astron.\ J.\  {\bf 116}, 1009 (1998);\
P.~Astier {\it et al.}  [The SNLS Collaboration],
Astron.\ Astrophys.\  {\bf 447}, 31 (2006);\
A.~G.~Riess {\it et al.},
Astrophys.\ J.\  {\bf 659}, 98 (2007).







\bibitem{Chiba:1999ka}
  T.~Chiba, T.~Okabe and M.~Yamaguchi,
  Phys.\ Rev.\  D {\bf 62}, 023511 (2000)
  [arXiv:astro-ph/9912463].

\bibitem{ArmendarizPicon:2000dh}
  C.~Armendariz-Picon, V.~F.~Mukhanov and P.~J.~Steinhardt,
  Phys.\ Rev.\ Lett.\  {\bf 85}, 4438 (2000)
  [arXiv:astro-ph/0004134].

\bibitem{ArmendarizPicon:2000ah}
  C.~Armendariz-Picon, V.~F.~Mukhanov and P.~J.~Steinhardt,
  Phys.\ Rev.\  D {\bf 63}, 103510 (2001)
  [arXiv:astro-ph/0006373].


\bibitem{ArmendarizPicon:1999rj}
  C.~Armendariz-Picon, T.~Damour and V.~F.~Mukhanov,
  Phys.\ Lett.\  B {\bf 458}, 209 (1999)
  [arXiv:hep-th/9904075].

\bibitem{Garriga:1999vw}
  J.~Garriga and V.~F.~Mukhanov,
  Phys.\ Lett.\  B {\bf 458}, 219 (1999)
  [arXiv:hep-th/9904176].


\bibitem{Sen:2002nu}
  A.~Sen,
  JHEP {\bf 0204}, 048 (2002)
  [arXiv:hep-th/0203211].

\bibitem{Sen:2002an}
  A.~Sen,
  Mod.\ Phys.\ Lett.\  A {\bf 17}, 1797 (2002)
  [arXiv:hep-th/0204143].

\bibitem{Gibbons:2002md}
  G.~W.~Gibbons,
  Phys.\ Lett.\  B {\bf 537}, 1 (2002)
  [arXiv:hep-th/0204008].

\bibitem{Bagla:2002yn}
  J.~S.~Bagla, H.~K.~Jassal and T.~Padmanabhan,
  Phys.\ Rev.\  D {\bf 67}, 063504 (2003)
  [arXiv:astro-ph/0212198].


\bibitem{Peebles:1987ek}
  P.~J.~E.~Peebles and B.~Ratra,
  Astrophys.\ J.\  {\bf 325}, L17 (1988).

\bibitem{Ratra:1987rm}
  B.~Ratra and P.~J.~E.~Peebles,
  Phys.\ Rev.\  D {\bf 37}, 3406 (1988).

\bibitem{Chiba:1997ej}
  T.~Chiba, N.~Sugiyama and T.~Nakamura,
  Mon.\ Not.\ Roy.\ Astron.\ Soc.\  {\bf 289}, L5 (1997)
  [arXiv:astro-ph/9704199].

\bibitem{Zlatev:1998tr}
  I.~Zlatev, L.~M.~Wang and P.~J.~Steinhardt,
  Phys.\ Rev.\ Lett.\  {\bf 82}, 896 (1999)
  [arXiv:astro-ph/9807002].


\bibitem{ArkaniHamed:2003uy}
  N.~Arkani-Hamed, H.~C.~Cheng, M.~A.~Luty and S.~Mukohyama,
  JHEP {\bf 0405}, 074 (2004)
  [arXiv:hep-th/0312099].

\bibitem{ArkaniHamed:2003uz}
  N.~Arkani-Hamed, P.~Creminelli, S.~Mukohyama and M.~Zaldarriaga,
  JCAP {\bf 0404}, 001 (2004)
  [arXiv:hep-th/0312100].

\bibitem{Nojiri:2005pu}
  S.~Nojiri and S.~D.~Odintsov,
  Gen.\ Rel.\ Grav.\  {\bf 38}, 1285 (2006)
  [arXiv:hep-th/0506212].

\bibitem{Capozziello:2005tf}
  S.~Capozziello, S.~Nojiri and S.~D.~Odintsov,
  Phys.\ Lett.\  B {\bf 632}, 597 (2006)
  [arXiv:hep-th/0507182].


\bibitem{Nojiri:2006be}
  S.~Nojiri and S.~D.~Odintsov,
  J.\ Phys.\ Conf.\ Ser.\  {\bf 66}, 012005 (2007)
  [arXiv:hep-th/0611071].


\bibitem{reconstruction}
S.~Nojiri, arXiv:0912.5066. 







\end{thebibliography}
\end{document}